\documentclass{ws-procs9x6}

\newcommand{\nn}{\nonumber}
\newcommand{\be}{\begin{equation}}
\newcommand{\ee}{\end{equation}}
\newcommand{\ba}{\begin{eqnarray}}
\newcommand{\ea}{\end{eqnarray}}

\def\gev{~{\rm GeV}}

\def\als{\alpha_{\rm s}}

\newcommand{\lsim}{\raisebox{-4pt}{$\,\stackrel{\textstyle
                                                         <}{\sim}\,$}}
\newcommand{\gsim}{\raisebox{-4pt}{$\,\stackrel{\textstyle
                                                         >}{\sim}\,$}}

\newcommand{\ov}[1]{\overline#1} 

\newcommand{\req}[1]{(\ref{#1})}

\def\xb{\bar{x}}
\def\sh{{s}}
\def\uh{{u}}
\begin{document}
\title{The handbag mechanism in wide-angle exclusive reactions}

\author{P.\ Kroll}

\address{Fachbereich Physik, Universit\"at Wuppertal,\\ 
D-42097 Wuppertal, Germany\\
Email: kroll@physik.uni-wuppertal.de}
\maketitle
\abstracts{
The handbag mechanism for wide-angle exclusive scattering reactions is 
discussed and compared with other theoretical approaches. Its application 
to  Compton scattering, meson photoproduction and two-photon annihilations 
into pairs of hadrons is reviewed in some detail.} 
\section{Introduction}
\begin{figure}[t]
\begin{center}
\includegraphics[width=4.0cm,bbllx=120pt,bblly=570pt,bburx=265pt,
bbury=680pt,clip=true]{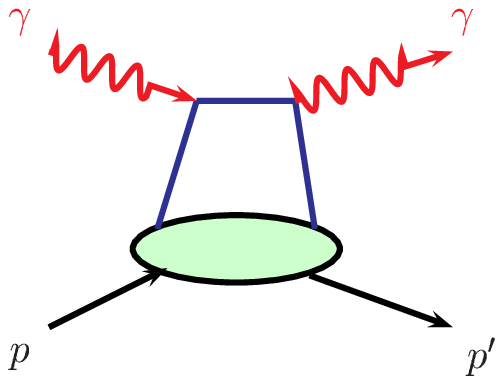} 
\includegraphics[width=4.0cm,bbllx=340pt,bblly=580pt,bburx=485pt,
bbury=693pt,clip=true]{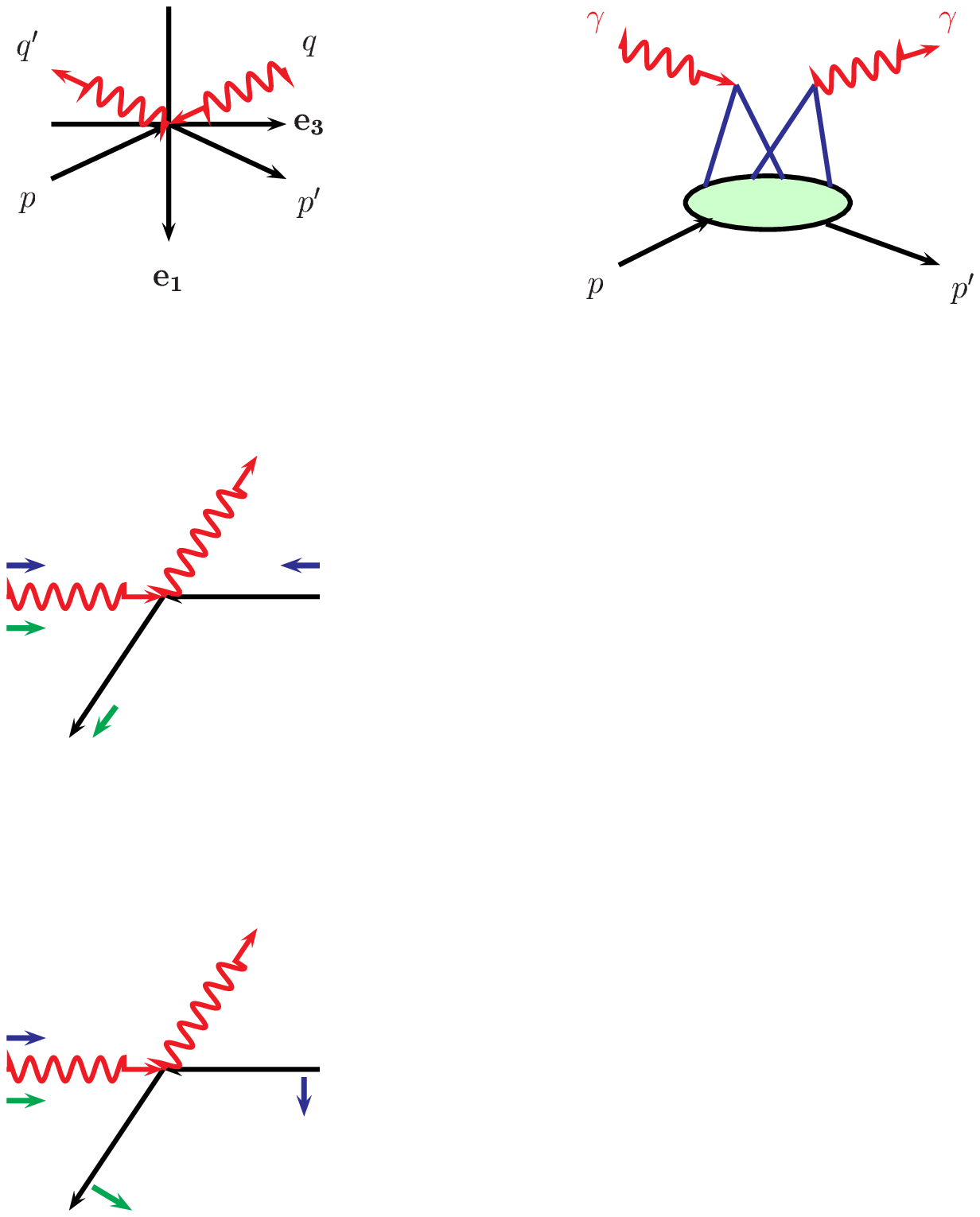} 
\includegraphics[width=5.1cm,bbllx=105pt,bblly=495pt,bburx=385pt,
bbury=610pt,clip=true]{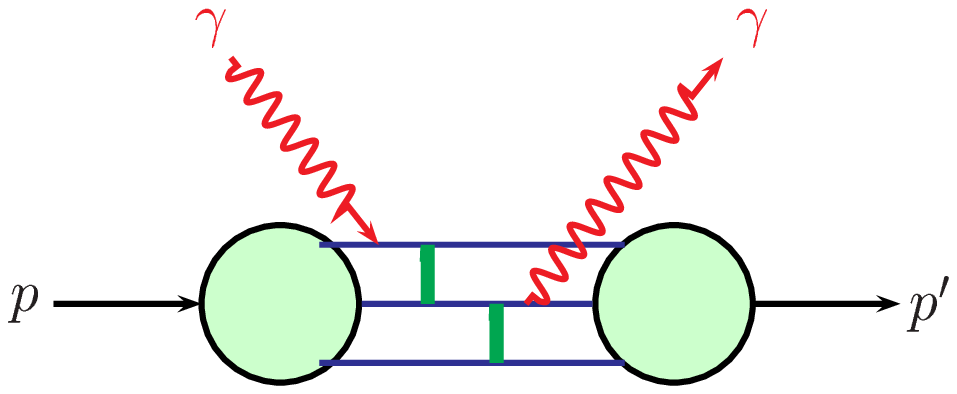} 
\caption{{}Handbag diagram for Compton scattering (upper left), cat's 
ears (upper right), and a leading-twist graph (lower left).}
\label{fig:handbag}
\end{center}
\end{figure}

Recently a new approach to wide-angle Compton scattering off protons
has been proposed \cite{rad98,DFJK1} where, for Mandelstam variables 
$s,-t,-u$ that are large as compared to a typical hadronic scale, $\Lambda^2$ 
of  the order of $1\gev^2$, the process amplitudes factorize into a hard 
parton-level subprocess, Compton scattering off quarks, and in soft form 
factors which represent $1/x$ moments of generalized parton distributions 
(GPDs) and encode the soft physics (see Fig.\ \ref{fig:handbag}). 
Subsequently it has been realized that this so-called handbag mechanism 
also applies to a number of other wide-angle reactions such as virtual 
Compton scattering \cite{DFJK2} (provided the photon virtuality, $Q^2$ is 
smaller than $-t$), meson photo- and electroproduction \cite{hanwen} 
or two-photon annihilations into pairs of mesons \cite{DKV2} or baryons 
\cite{DKV2,weiss}. It should be noted that the handbag mechanism bears 
resemblance to the treatment of inelastic Compton scattering advocated for 
by Bjorken and Paschos \cite{bjo} long time ago.  

There are other mechanisms which also contribute to wide-angle scattering 
besides the handbag which is characterized by one active parton, 
i.e.\ one parton from each hadron participates in the hard subprocess
(e.g.\ $\gamma q\to \gamma q$ in Compton scattering) while all others are
spectators. On the one hand, there are the so-called cat's ears graphs 
(see Fig.\ \ref{fig:handbag}) with two active partons participating in the 
subprocess (e.g.\ $\gamma qq \to \gamma qq$). It can be shown however that 
in these graphs either a large parton virtuality or a large parton 
transverse momentum occurs. This forces the exchange of at least one hard 
gluon. Hence, the cat's ears contribution is expected to be suppressed 
as compared to the handbag one. The next class of graphs are characterized 
by three active quarks (e.g.\ $\gamma qqq \to \gamma qqq$) and, obviously, 
require the exchange of at least two hard gluons. For, say, Compton 
scattering off protons, the so-called leading-twist contribution (see Fig.\ 
\ref{fig:handbag}) for which all valence quarks participate in the hard 
process, belong to this class\cite{bro80}. The leading-twist factorization 
is given by a convolution of the hard subprocess, e.g.\ 
$\gamma qqq \to \gamma qqq$ in Compton scattering off protons and 
distribution amplitudes encoding the soft physics. This contribution is 
expected to dominate for asymptotically large momentum transfer~\footnote{
Interestingly, for the pion-photon transition form factor the 
handbag and the leading-twist contributions fall together}. 
Formally, the handbag contribution is a power correction to the 
leading-twist one.

Since hadrons are not just made off their valence quarks one go on and 
consider four active partons and so forth. The series generated that way,
bears resemblance to an expansion in terms of n-body operators used 
in many-body theory. In principle, all the different contributions have 
to be added coherently. In practice, however, this is a difficult, 
currently almost impossible task~\footnote{
An exception is the pion's electromagnetic form factors where this has 
been attempted by several groups, see for instance \cite{JKR}} 
since each contribution has its own associated soft hadronic matrix 
element which, as yet, cannot be calculated from QCD and is often even 
phenomenologically unknown. We have to learn from experiment, 
presently characterized by momentum transfers of the order of 
$10 \gev^2$, whether one of the mentioned mechanisms is dominant  
or whether the coherent sum of some or all topologies is 
actually needed. 
 
The handbag mechanism in real Compton scatering is reviewed in some 
detail in Sect.\ 2. The large $-t$ behaviour of the GPDs and their 
associated form factors is discussed in Sect.\ 3 and predictions for 
Compton scattering are given. A few results for wide-angle meson 
photoproduction and two-photon annihilations into pairs of hadrons  
are presented in Sect.\ 4 and 5, respectively. The paper ends with a 
summary (Sect.\ 6).   
\section{Wide-angle Compton scattering}
For Mandelstam variables $s$, $-t$ and $-u$ that are large as compared
to a typical hadronic scale $\Lambda^2$ where $\Lambda$ being of order
$1\, \gev$, it can be shown that the handbag diagram shown in
Fig.\ \ref{fig:handbag}, is of relevance in wide-angle Compton scattering. 
To see this it is of advantage to work in a symmetrical frame which is 
a c.m.s rotated in such a way that the momenta of the incoming ($p$) 
and outgoing ($p'$) proton momenta have the same light-cone plus 
components. In this frame the skewness, defined as 
\be 
\xi = \frac{(p - p')^+}{(p + p')^+}\,,
\ee
is zero. The bubble in the handbag is viewed as a sum over all possible 
parton configurations as in deep ineleastic lepton-proton scattering. 
The crucial assumptions in the handbag approach are that of restricted
parton virtualities, $k_i^2<\Lambda^2$, and of intrinsic transverse
parton momenta, ${\bf k_{\perp i}}$, defined with respect to their
parent hadron's momentum, which satisfy $k_{\perp i}^2/x_i
<\Lambda^2$, where $x_i$ is the momentum  fraction parton $i$ carries.   
 
\begin{figure}[t]
\begin{center}
\includegraphics[width=3.3cm,bbllx=261pt,bblly=606pt,bburx=380pt, 
bbury=700pt,clip=true]{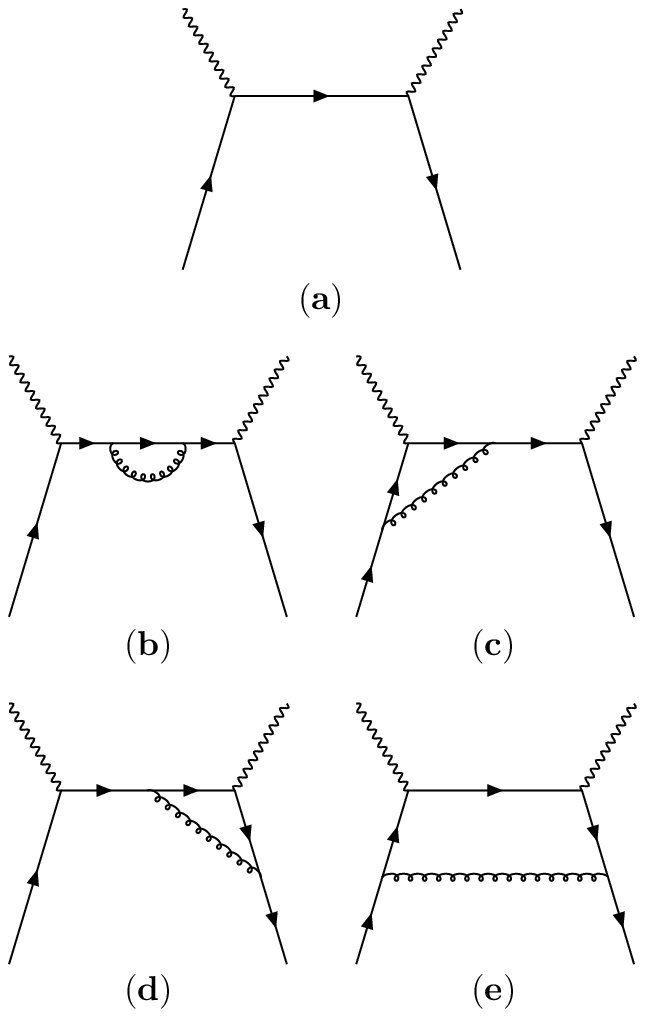}  
\includegraphics[width=4.3cm,bbllx=212pt,bblly=405pt,bburx=400pt, 
bbury=600pt,clip=true]{handbag-NLO.ps} 
\caption{{}Sample LO (a) and NLO (b-e) pQCD Feynman graphs for the 
partonic subprocess, $\gamma q\to \gamma q$, in the handbag mechanism.} 
\label{fig:LONLO}
\end{center}
\end{figure}

One can then show \cite{DFJK1} that the subprocess Mandelstam variables
$\hat{s}$ and $\hat{u}$ are the same as the ones for the full process,
Compton scattering off protons, up to corrections of order
$\Lambda^2/t$:
\ba
\hat{s}=(k_j+q)^2 \simeq (p+q)^2 =s\,, \quad 
\hat{u}=(k_j-q')^2 \simeq (p-q')^2 =u\,.
\ea
The active partons, i.e.\ the ones to which the photons couple, are
approximately on-shell, move collinear with their parent hadrons and
carry a momentum fraction close to unity, $x_j, x_j' \simeq 1$.
Thus, like in deep virtual Compton scattering, the physical 
situation is that of a hard parton-level subprocess, 
$\gamma q\to \gamma q$, and a soft emission and reabsorption of quarks 
from the proton. The light-cone helicity amplitudes \cite{diehl01} for 
wide-angle Compton scattering then read
\ba
{M}_{\mu'+,\,\mu +}(s,t) &=& \; \frac{e^2}{2}\,
     \left[\, { T}_{\mu'+,\,\mu+}(\sh,t)\,(R_V(t) + R_A(t))\,\right.\nn\\[0.3em]
&&\qquad\left.  + \,\,  { T}_{\mu'-,\,\mu-}(\sh,t)\,(R_V(t) - R_A(t)) \right]  
                                                 \,, \label{ampl}\\[0.5em]
 { M}_{\mu'-,\,\mu +}(s,t) &=& \; \frac{e^2}{2}\, \frac{\sqrt{-t}}{2m} 
         \left[\,  T_{\mu'+,\,\mu+}(\sh,t)\, 
         + \,  { T}_{\mu'-,\,\mu-}(\sh,t)\, \right] \,R_T(t)\,.\nn
\label{eq:amp}
\ea
$\mu,\, \mu'$ denote the helicities of the incoming and outgoing
photons, respectively. The helicities of the protons in $ { M}$ and
of the quarks in the hard scattering amplitude $ T$ are labeled by their
signs. $m$ denotes the mass of the proton. The form factors $R_i$ 
represent $1/\xb$-moments of GPDs at zero skewness. This 
representation which requires the dominance of the plus components of 
the proton matrix elements, is a non-trivial feature given that, in 
contrast to deep inelastic lepton-nucleon and deep virtual Compton 
scattering, not only the plus components of the proton 
momenta but also their minus and transverse components are large here. 
The hard scattering has been calculated to next-to-leading order (NLO) 
perturbative QCD \cite{HKM}, see Fig.\ \ref{fig:LONLO}. It turned out 
that the NLO amplitudes are ultraviolet regular but those amplitudes 
which are non-zero to LO, are infrared divergent. As usual the infrared
divergent pieces are interpreted as non-perturbative physics and absorbed
into the soft form factors, $R_i$. Thus, factorization of the wide-angle
Comton amplitudes within the handbag approach is justified to (at least)
NLO. To this order the gluonic subprocess, $\gamma g\to \gamma g$, has 
to be taken into account as well which goes along with corresponding 
gluonic GPDs and their associated form factors.

The handbag amplitudes \req{ampl} lead to the following result for the
Compton cross section  
\ba
\frac{d\sigma}{dt} &=& \frac{d\hat{\sigma}}{dt} \left\{ \frac12\, \big[
R_V^2(t)\,(1+\kappa_T^2) + R_A^2(t)\big] \right.\nn\\
&&\left. \quad  - \frac{\uh\sh}{\sh^2+\uh^2}\, \big[R_V^2(t)\,(1+\kappa_T^2) 
                - R_A^2(t)\big]\right\} + O(\alpha_s)\,,
\label{dsdt}
\ea
where $d\hat{\sigma}/dt$ is the Klein-Nishina cross section for
Compton scattering off massless, point-like spin-1/2 particles of
charge unity. The parameter $\kappa_T$ is defined as 
\be
\kappa_T= \frac{\sqrt{-t}}{2m}\, \frac{R_T}{R_V}\,. 
\ee
Another interesting observable in Compton scattering is the helicity
correlation, $A_{LL}$,  between the initial state photon and proton
or, equivalently, the helicity transfer, $K_{LL}$, from the incoming
photon to the outgoing proton. In the handbag approach one obtains
\cite{DFJK2,HKM} 
\be
A_{LL}=K_{LL}\simeq \frac{\sh^2 - \uh^2}{\sh^2 + \uh^2}\, 
                    \frac{R_A}{R_V} + O(\kappa_T,\alpha_s)\,,
\label{all}
\ee  
where the factor in front of the form factors is the corresponding
observable for $\gamma q\to \gamma q$. The result \req{all} is a
robust prediction of the handbag mechanism, the magnitude of the
subprocess helicity correlation is only diluted  somewhat by the 
ratio of the form factors $R_A$ and $R_V$. 
\section{The large-$t$ behaviour of GPDs}
\label{sect:model}
In order to make actual predictions for Compton scattering a model for 
the form factors or rather for the underlying GPDs is required.
A first attempt to parameterize the GPDs $H$ and $\widetilde{H}$  
at zero skewness is~\cite{rad98,DFJK1,HKM} 
\ba
H^a(\xb,0;t) &=& \exp{\left[a^2 t
        \frac{1-\xb}{2\xb}\right]}\, q_a(\xb)\,,\nn\\ 
\widetilde{H}^a(\xb,0;t) &=& \exp{\left[a^2 t
        \frac{1-\xb}{2\xb}\right]}\, \Delta q_a(\xb)\,,
\label{gpd}
\ea
where $q(\xb)$ and $\Delta q(\xb)$ are the usual unpolarized and 
polarized parton distributions in the proton~\footnote
{The parameterization \req{gpd} can be motivated by overlaps of 
light-cone wave functions which have a Gaussian $\vec{k}_\perp$ 
dependence \cite{rad98,DFJK1,DFJK3}.}. 
The transverse size of the proton, $a$, is the only free parameter 
and even it is restricted to the range of about 0.8 to 1.2 $\gev^{-1}$. 
Note that $a$ essentially refers to the lowest Fock states of the 
proton which, as phenomenological experience tells us, are rather 
compact. The model (\ref{gpd}) is designed for large $-t$. Hence, 
forced by the Gaussian in (\ref{gpd}), large $\xb$ is implied, too. 
Despite of this the normalizations of the model GPDs at $t=0$ are correct.
Since the phenomenological parton distributions, see e.g.\ \cite{GRV}, 
suffer from large uncertainties at large $x$, the GPDs (\ref{gpd}) have
been improved in \cite{DFJK1} by using overlaps of light-cone wave 
functions for $x \gsim 0.6$  instead of the GRV 
parameterization \cite{GRV}.    

\begin{figure}[t]
\begin{minipage}{0.46\textwidth}
\begin{center} 
\includegraphics[width=4.0cm,bbllx=90pt,bblly=30pt,bburx=590pt,
bbury=635pt,angle=-90,clip=true]{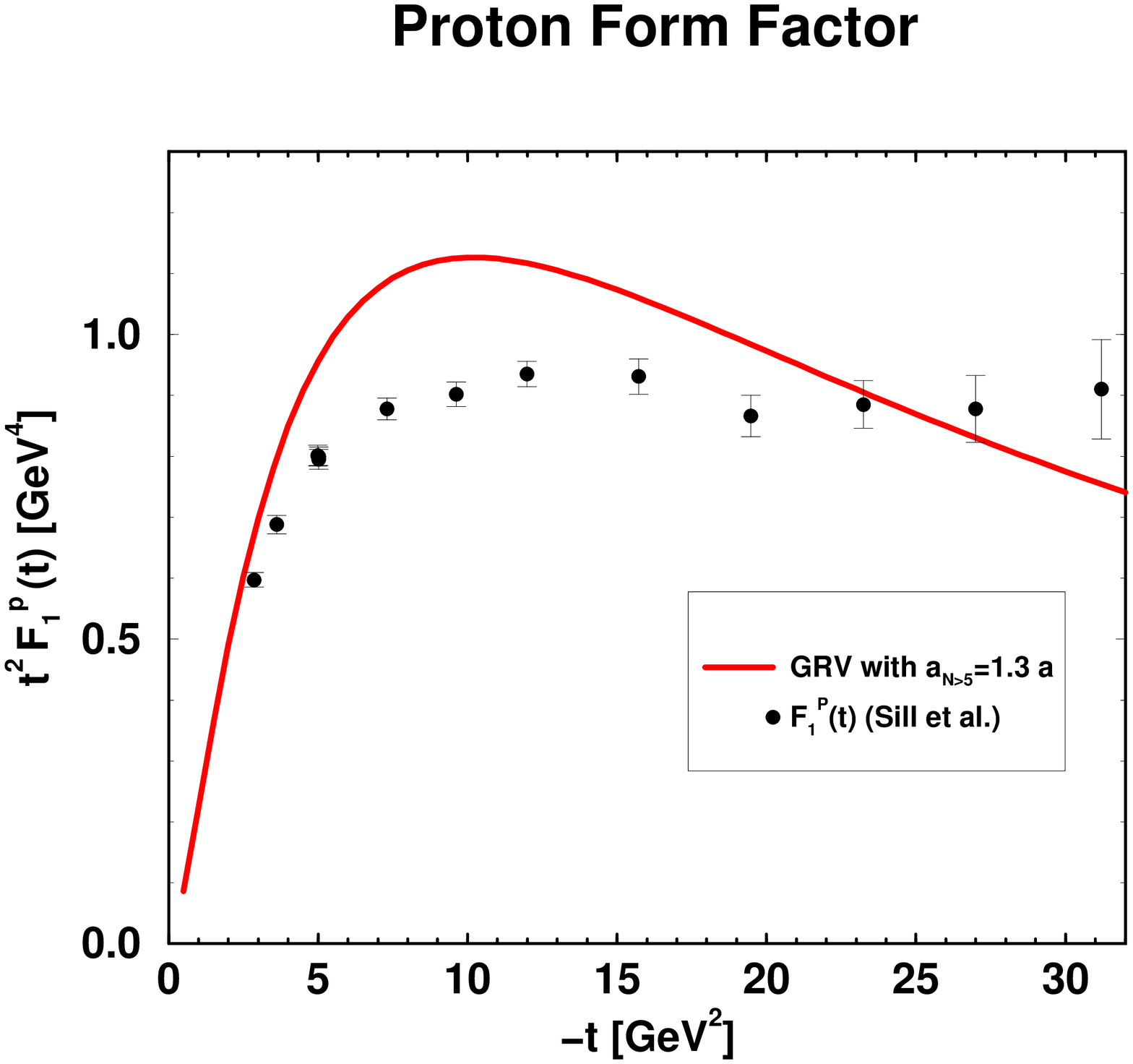}
\end{center}
\end{minipage}
\begin{minipage}{0.46\textwidth}
\begin{center}
\includegraphics[width=4.9cm,bbllx=50pt,bblly=107pt,bburx=398pt, 
bbury=390pt,clip=true]{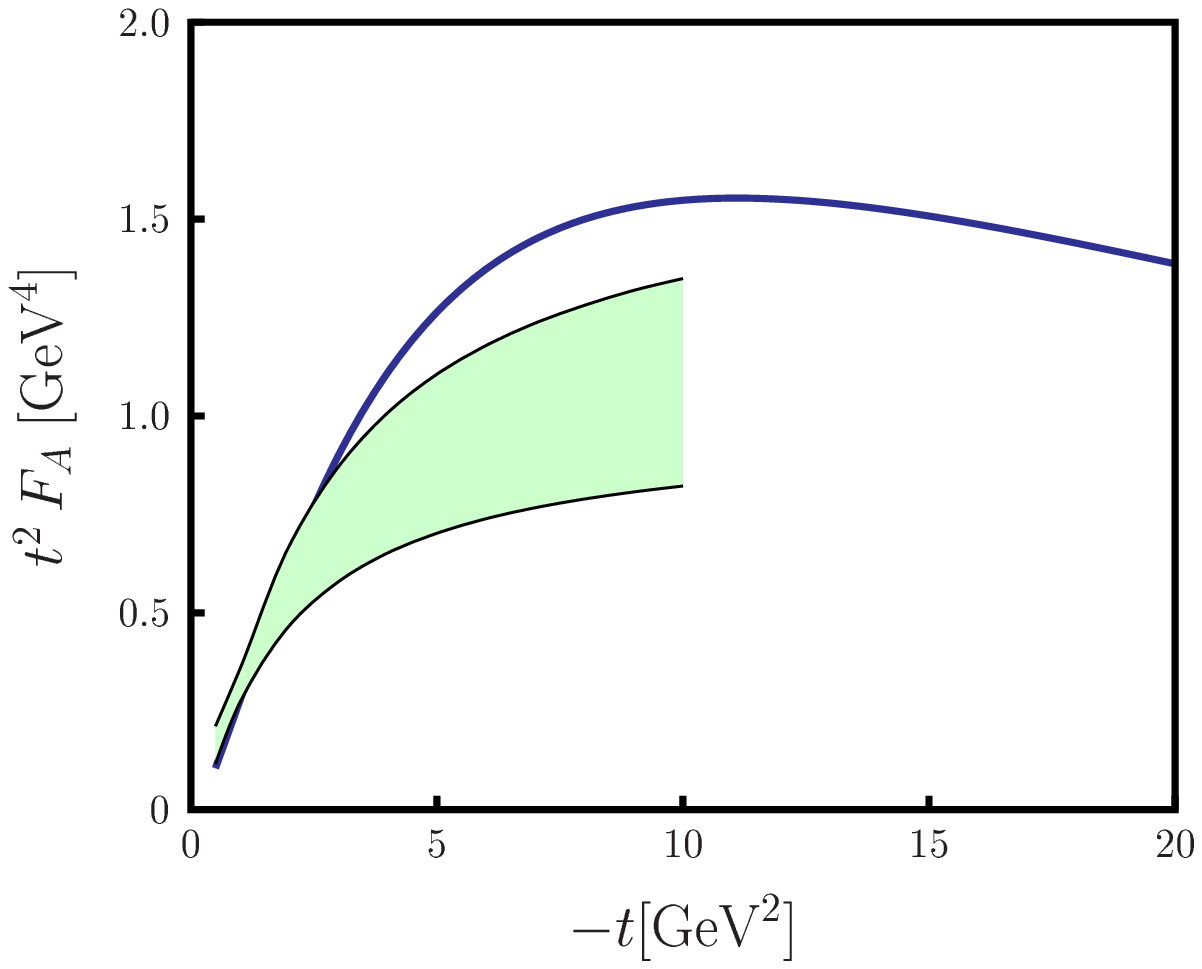}
\end{center}
\end{minipage} 
\caption{The Dirac form factor of the proton (left) and the axial 
vector form factor (right) scaled by $t^2$, are plotted vs. $t$. 
Data are taken from Ref.\ \protect\cite{sill}. The band represents 
a dipole fit to the neutrino data \protect\cite{kita}. 
The theoretical results are taken from \protect\cite{DFJK1}.} 
\label{fig:form}
\end{figure}

\begin{figure}[t]
\begin{center}
\includegraphics[width=5.8cm,bbllx=27pt,bblly=47pt,bburx=398pt, 
bbury=295pt,clip=true]{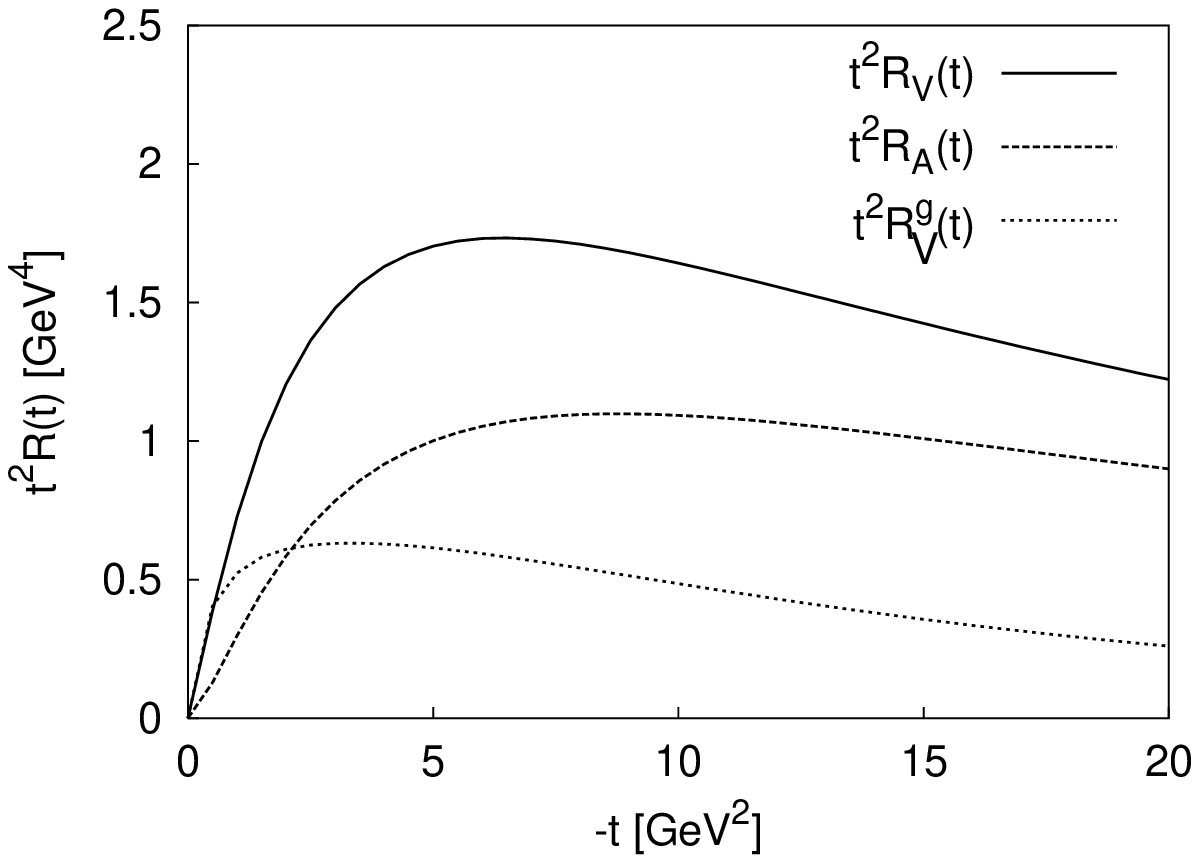}
\includegraphics[width=5.5cm,bbllx=50pt,bblly=50pt,bburx=400pt, 
bbury=300pt,clip=true]{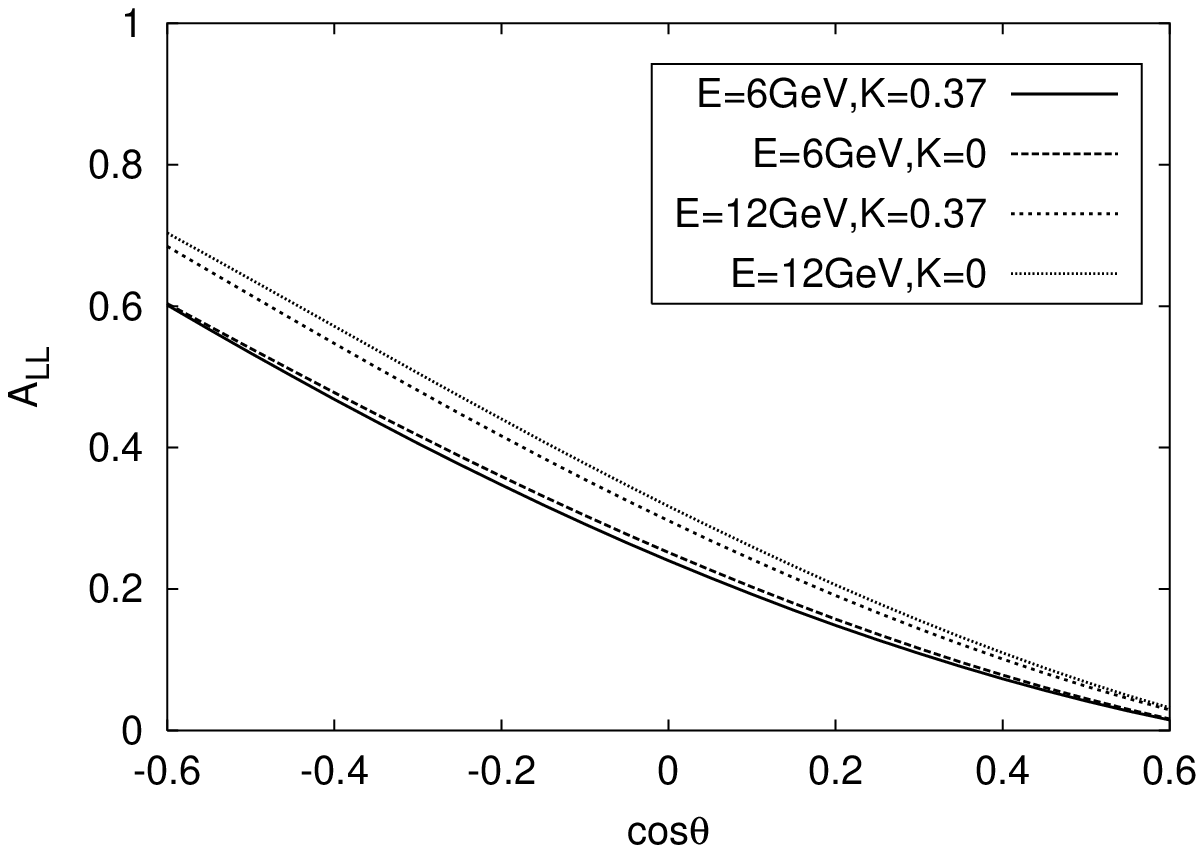} 
\caption{Predictions for the Compton form factors (left) and for 
the helicity correlations $A_{LL}=K_{LL}$ (right). NLO
corrections and the tensor form factor are taken into account 
(scenario A), in scenario B they are neglected.} 
\label{fig:cross}
\end{center}
\end{figure}

With the model GPDs \req{gpd} at hand one can evaluate the various form
factors by taking appropriate moments. For the Dirac and the axial form 
factor one has
\be
F_1=\sum_q e_q \int_{-1}^1 d\xb H^q(\xb,0;t),\quad
F_A= \int_{-1}^1 d\xb \Big[ \widetilde{H}^u(\xb,0;t)-
             \widetilde{H}^d(\xb,0;t)\Big],
\label{formfactors}
\ee
while the Compton form factors read
\be
R_V=\sum_q e_q^2 \int_{-1}^1 \frac{d\xb}{\xb} H^q(\xb,0;t), \quad
R_A=\sum_q e_q^2 \int_{-1}^1 \frac{d\xb}{\xb} {\rm sign}(\xb)\, 
\widetilde{H}^q(\xb,0;t).
\label{Compton-formfactors}
\ee
Results for the nucleon form factors are shown in Fig.\ \ref{fig:form}. 
Obviously, as the comparison with experiment \cite{sill,kita} reveals, 
the model GPDs work quite well although the predictions for the Dirac 
and the axial form factors overshoot the data by about 
$20 - 30\,\%$ for $-t$ around $5 \gev^2$. An effect of similar size can 
be expected for the Compton form factors for which predictions are 
shown in Fig.\ \ref{fig:cross}. The scaled form factors $t^2 F_{1,A}$ 
and $t^2 R_i$ exhibit broad maxima which mimick dimensional counting 
in a range of $-t$ from, say, $5$ to about $20\,\gev^2$. The position 
of the maximum of any of the scaled form factors is approximately 
located at \cite{DFJK2} 
\be
t_0 \simeq -4 a^{-2}\, \left\langle
                          \frac{1-\xb}{\xb}\right\rangle^{-1}_{F(R)}\,.
\label{max-pos}
\ee
The mildly $t$-dependent mean value $\langle (1-\xb)/\xb\rangle$ comes
out around $1/2$. A change of $a$ moves the position of the maximum of 
the scaled form factors but leaves their magnitudes essentially unchanged.
It is tempting to assume that form factors of the type discussed here 
also control other wide-angle reactions as, for instance, elastic 
hadron-hadron scattering~\cite{DFJK2}~\footnote{
This is similar to the parton scattering model discussed 30 years 
ago, see e.g.\ \cite{const}}. The experimentally observed approximate
scaling behaviour of these cross sections is then attributed to the 
broad maxima the scaled form factors show. I.e.\ the scaling behaviour 
observed for momentum transfers of the order of $10 \gev^2$, reflects
rather the transverse size of the hadrons \req{max-pos} than a property 
of the leading-twist contribution~\footnote{
The apparent absence of perturbative logs generated by the running of 
$\als$ and the evolution of the distribution amplitudes and which are
characteristic of a perturbative calculation, is a clear signal against 
the latter interpretation}.

The Pauli form factor $F_2$ and its Compton analogue $R_T$ contribute
to proton helicity flip matrix elements and are related to the GPD $E$
analogously to \req{formfactors}. This connection suggests that, at
least for not too small values of $-t$, $R_T/R_V$ roughly behaves as
$F_2/F_1$. Thus, the recent JLab data~\cite{gayou} on $F_2$ indicate 
a behaviour as $R_T/R_V\propto m/\sqrt{-t}$. The form factor $R_T$ 
therefore contributes to the same order in $\Lambda/\sqrt{-t}$ as 
the other ones, see \req{eq:amp}. Predictions for Compton observables 
are given for two different scenarios~\footnote{
There is a discrepancy between the SLAC data~\cite{slac} on 
$F_2/F_1$, obtained by Rosenbluth separation, and the JLab ones. 
According to Ref.\ \cite{tjon}, part of the discrepancy can be assigned 
to two-photon exchange which affects the Rosenbluth data~\cite{slac}}. 
Both $R_T$ and $\als$ corrections are omitted in scenario B but taken 
into account in A where the ratio $\kappa_T$ is assumed to have a value 
of 0.37 as estimated from the JLab form factor data~\cite{gayou}.


Employing the model GPDs and the corresponding form factors, various
Compton observables can be calculated \cite{DFJK1,DFJK2,HKM}. The
predictions for the differential cross section are in fair agreement
with the Cornell data~\cite{shupe}. Due to the broad maxima the 
scaled form factors exhibit, the handbag mechanism approximately predicts
a $s^6$-scaling behaviour at fixed c.m.s. scattering angle according 
to dimensional counting. Closer inspection of the handbag predictions 
however reveals that the effective power of $s$ depends on the scattering 
angle and on the range of energy used in the determination of the power.  
The JLab E99-114 collaboration \cite{nathan} will provide accurate 
cross section data soon which will allow for a crucial examination of the 
handbag mechanism and may necessitate an improvement of the model GPDs 
(\ref{gpd}).   

Predictions for $A_{LL}=K_{LL}$ are shown in Fig.\ \ref{fig:cross}.
The JLab E99-114 collaboration \cite{nathan} has presented a first
measurement of $K_{LL}$ at a c.m.s.\ scattering angle of
$120^\circ$ and a photon energy of $3.23 \gev$. This still
preliminary data point is in fair agreement with the predictions
from the handbag given the small energy at which they are
available. The kinematical requirement of the handbag mechanism 
$s,\; -t,\; -u \gg \Lambda^2$ is not well satisfied and therefore one
has to be aware of large dynamical and kinematical corrections 
(proton mass effects have been investigated in Ref.\ \cite{DFHK}). 

In the introduction I mentioned the leading-twist factorization 
scheme~\cite{bro80} for which all valence quarks of the involved hadrons 
participate in the hard scattering and not just a single one. The 
leading-twist calculations, e.g.\ \cite{dixon}, reveal difficulties 
in getting the size of the Compton cross section correctly, the 
numerical results are way below experiment. There is growing 
evidence~\cite{bolz}~\footnote{
A perturbatively calculated $J/\Psi\to p\bar{p}$ decay width only agrees
with experiment if a proton distribution amplitude close to the
asymptotic form is employed~\cite{bolz96}}
that the proton's leading-twist distribution 
amplitude is close to the asymtotic form $\propto x_1 x_2 x_3$. Using 
such a distribution amplitude in a leading-twist calculation of the 
Compton cross section, the result turns out to be too small by a 
factor of about $10^{-3}$. Moreover, the leading-twist 
approach~\cite{dixon} leads to a negative value for $K_{LL}$ at
angles larger than $90^\circ$ in conflict with the JLab 
result~\cite{nathan}. Thus, we are forced to conclude that wide-angle
Compton scattering at energies available at JLab is not dominated by the 
leading-twist contribution.     

The handbag approach to real Compton scattering can straightforwardly  
be extended to virtual Compton scattering \cite{DFJK2} provided 
$Q^2/-t \lsim 1$. Recently, the NLO corrections to the hard subprocess
have been calculated for virtual Compton scattering \cite{morii}.  
\section{Meson photoproduction}
Photo- and electroproduction of mesons have also been discussed
within the handbag approach~\cite{hanwen} using, as in deep virtual
electroproduction~\cite{dvem}, a one-gluon exchange mechanism for the
generation of the meson. As it turns out the one-gluon exchange 
contribution fails with the normalization of the photoproduction
cross section by order of magnitude. Either vector meson dominance
contributions are still large or the generation of the meson
by the exchange of a hard gluon underestimates the handbag
contribution. Since the same Feynman graphs contribute here as in the case 
of the pion's electromagnetic form factor the failure of the one-gluon
exchange contribution is perhaps not a surprise~\cite{JKR}. 

One may investigate the handbag contribution to photoproduction of 
pseudoscalar mesons (P) in a more general way~\cite{HJKP} by writing 
down a covariant decomposition~\cite{CGLN} of the subprocess 
$\gamma q\to P q$ in terms of four covariants which take care of the 
helicity dependence in the subprocess, and four invariant functions which 
encode the dynamics. Assuming dominance of quark helicity non-flip 
one finds, for instance, that the helicity correlation $\hat{A}_{LL}$ 
for the subprocess $\gamma q\to P q$ is the same as for 
$\gamma q \to \gamma q$, see (\ref{all}). $A_{LL}$ for the full process 
is similar to the result \req{all} for Compton scattering, too. Another 
interesting result is the ratio of the cross sections for 
photoproduction of $\pi^+$ and $\pi^-$. The ratio is approximately given by
\be
\frac{d\sigma(\gamma n\to \pi^- p)}{d\sigma(\gamma p\to \pi^+ n)} \simeq
\left[\frac{e_d \uh + e_u \sh}{e_u \uh + e_d \sh}\right]^2\,.
\label{pi-ratio}
\ee
The form factors which, for a given flavor, are the same as those
appearing in Compton scattering, cancel in the ratio. The 
prediction~\req{pi-ratio} is
in fair agreement with a recent JLab measurement \cite{zhu} 
which, at $90^\circ$, provides values of $1.73\pm 0.15$ and $1.70\pm 0.20$
for the ratio at beam energies of $4.158$ and $5.536 \gev$, respectively.
This result supports the handbag mechanism with dominant
quark helicity non-flip.
\section{Two-photon annihilations into pairs of hadrons} 
The arguments for handbag factorization hold as well for two-photon 
annihilations into pairs of hadrons as has recently been shown in Ref.\ 
\cite{DKV2} (see also Ref.~\cite{weiss}). 
The cross section for the production of a pair of pseudoscalar mesons
reads 
\be
\frac{d\sigma}{dt}(\gamma\gamma\to M\ov{M}) = 
     \frac{8\pi\alpha^2_{\rm elm}} 
              {s^2 \sin^4 \theta} \big|R_{M\ov{M}}(s)\big|^2\,,
\ee
while for baryon pairs it is given by
\ba
\frac{d\sigma}{dt}\,(\,\gamma\gamma\,\to\,\, B\ov{B}\,) &=& 
      \frac{4\pi\alpha^2_{\rm elm}} 
        {s^2 \sin^2 \theta} \Big\{ \big|R_A^B(s)+ R_P^B(s)\big|^2\nn\\
                  &+& \cos^2\theta\,\big|R_V^B(s)\big|^2\,+\, 
                     \frac{s}{4m^2}\,\big| R_P^B(s)\big|^2 
                                       \Big\}\,.
\ea
In analogy to Eq.\ \req{Compton-formfactors} the form factors represent
moments of two-hadron distribution amplitudes, $\Phi_{2h}$, which are 
time-like versions of GPDs. In the case of pion pair production one 
has for instance 
\be
R_{2\pi}(s)=\sum_q e_q^2 R^q_{2\pi}(s)\,, \quad
      R^q_{2\pi}(s)=\frac12 \int_0^1 dz (2z-1) \Phi_{2\pi}(z,1/2,s)\,.
\ee

\begin{figure}[t]
\begin{center}
\includegraphics[width=5.0cm]{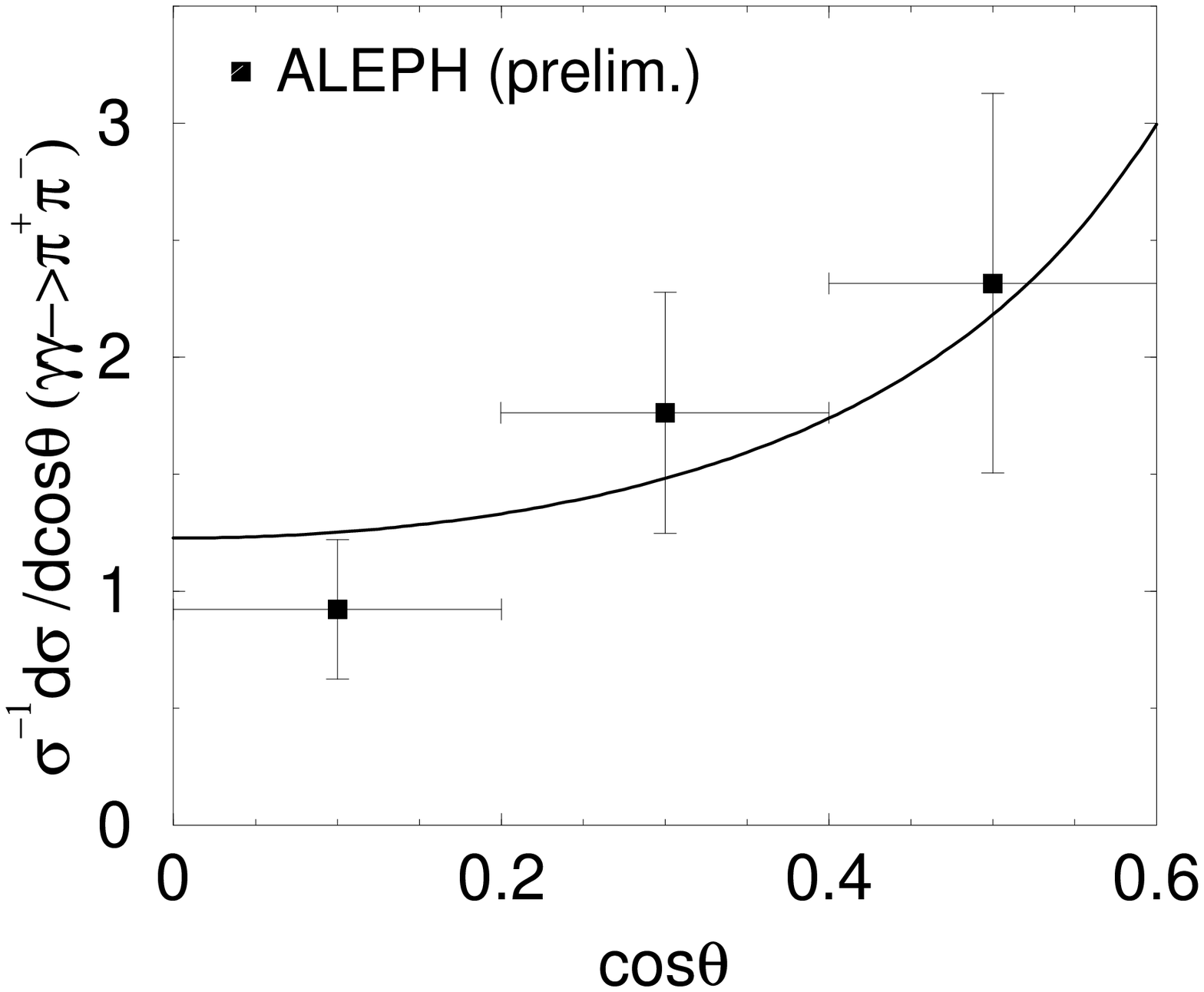}
\includegraphics[width=5.5cm]{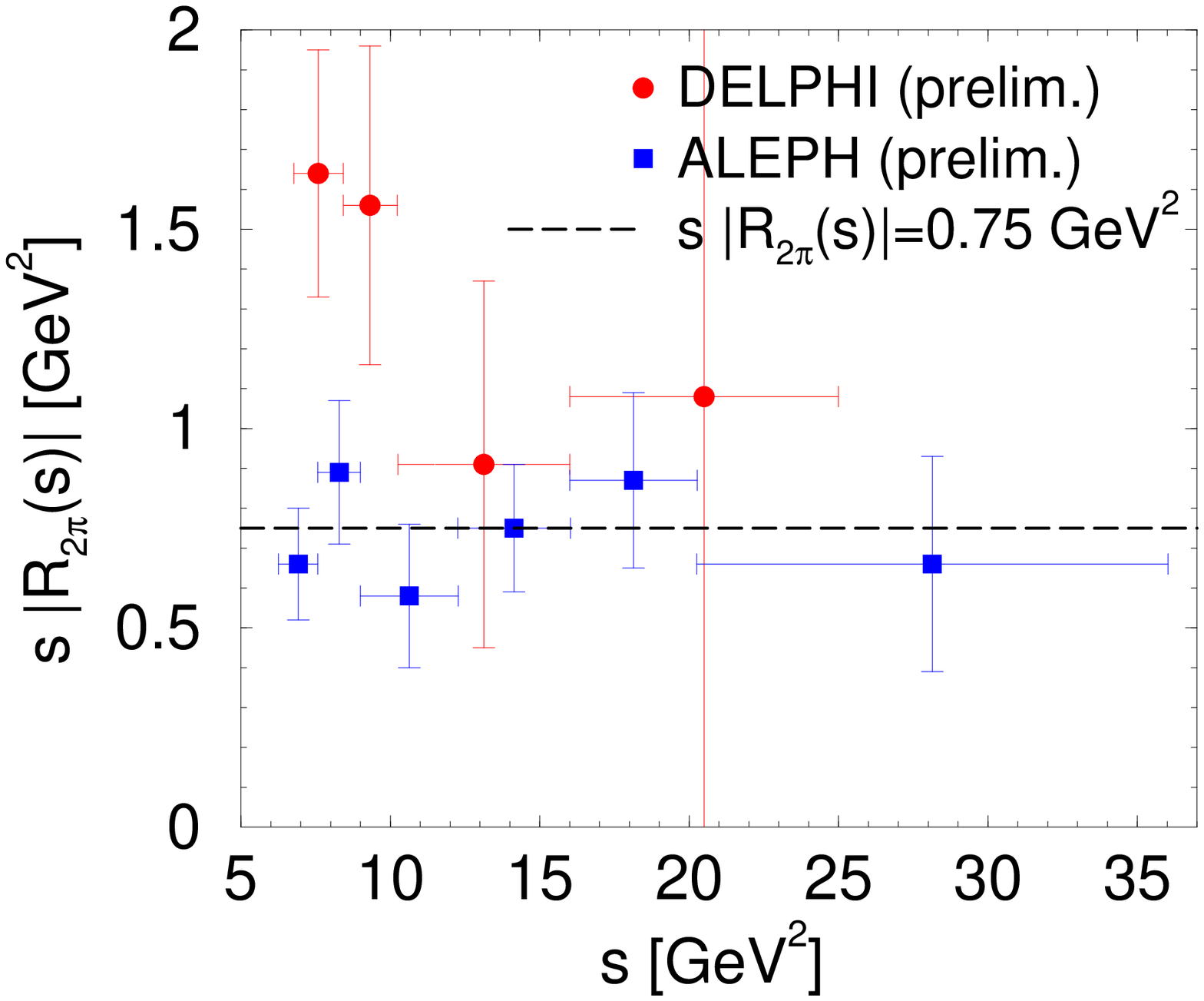}
\caption{Handbag predictions for the angular dependence of the cross section
for $\gamma\gamma\to \pi^+\pi^-$ (left) and the form factor $s|R_{2\pi}|$ 
versus $s$ (right). Preliminary data are taken from ALEPH~\protect\cite{ALEPH} 
and DELPHI~\protect\cite{DELPHI}.}
\label{fig:pipi}
\end{center}
\end{figure}

The angle dependencies of the cross sections which are (almost) 
independent of the form factors, are in fair agreement with experiment, 
see Fig.\ \ref{fig:pipi}. The form factors have not been modelled in 
Refs.\ \cite{DKV2} but rather extracted ('measured') from the 
experimental cross section. The form factor $R_{2\pi}$ obtained that way,
is shown in Fig.\ \ref{fig:pipi}, too. The average
value of the scaled form factor $sR_{2\pi}$ is $0.75\gev^2$. The closeness
of this value to that of the scaled time-like electromagnetic form factor 
of the pion ($0.93 \pm 0.12 \gev^2$) hints at the internal consistency 
of the handbag approach. 

A characterisic feature of the handbag mechanism in the time-like
region is the intermediate $q\ov{q}$ state implying the absence of 
isospin-two components in the final state. A consequence of this property is 
\be
\frac{d\sigma}{dt}(\gamma\gamma\to \pi^0\pi^0) = 
                 \frac{d\sigma}{dt}(\gamma\gamma\to \pi^+\pi^-)\,,
\ee
which is independent of the soft physics input and is, in so
far, a robust prediction of the handbag approach. 
The absence of the isospin-two components combined with flavor
symmetry allows one to calculate the cross sections for other $B\ov{B}$
channels using the form factors for $p\ov{p}$ as the only soft physics
input. It is to be stressed that the leading-twist mechanism has again 
difficulties to account for the size of the cross sections \cite{farrar}
while the diquark model \cite{schweiger} which is a variant of the 
leading-twist approach in which diquarks are considered as 
quasi-elementary constituents of baryons, is infair agreement with
experiment for $\gamma\gamma\to B\bar{B}$.
\section{Summary}
I have reviewed the theoretical activities on applications of the
handbag mechanism to wide-angle scattering. There are many interesting
predictions, some are in fair agreement with experiment, others still
awaiting their experimental examination. It seems that the handbag
mechanism plays an important role in exclusive scattering
for momentum transfers of the order of $10\gev^2$. However, before we
can draw firm conclusions more experimental tests are needed.
The leading-twist approach, on the other hand, typically provides
cross sections which are way below experiment. As is well-known the
cross section data for many hard exclusive processes exhibit
approximate dimensional counting rule behaviour. Infering
from this fact the dominance of the leading-twist contribution is
premature. The handbag mechanism can explain this approximate
power law behaviour (and often the magnitude of the cross sections),
too. It is attributed to the broad maxima the scaled form factors
show and, hence, reflects the the transverse size of the lowest Fock
states of the involved hadrons.  

I finally emphasize that the structure of the handbag amplitude,
namely its representation as a product of perturbatively calculable
hard scattering amplitudes and $t$-dependent form factors is the
essential result. Refuting the handbag approach necessitates  
experimental evidence against this factorization.

\end{document}